\documentclass[12pt, draftclsnofoot, onecolumn]{IEEEtran}
\usepackage[table]{xcolor}
 \usepackage{comment,graphicx}
 \usepackage{subfigure}
 \usepackage{epstopdf}
\usepackage{psfrag}
\usepackage{acronym}
\usepackage{etoolbox}
\usepackage{epsfig}
\usepackage{colortbl,color}
\usepackage{amsmath}
\usepackage{amssymb}
\usepackage{mathabx}
\usepackage{enumerate}
\usepackage{bbm}
\usepackage{algorithm}
\usepackage{algorithmic}
\usepackage{lipsum,multicol}
\usepackage{stfloats}
\usepackage{arydshln} 
\usepackage{amsfonts}
\usepackage{dsfont}
\usepackage{upgreek}
\usepackage{soul}
\usepackage{tabularx}
\usepackage{multirow}
\usepackage{graphicx}

\usepackage{cite}
\usepackage{verbatim}
\usepackage{blkarray}
\usepackage{bm}
\usepackage{tikz}
\usepackage{booktabs}

\newcommand{\alp}{\alpha_{nm}}

\newcommand{\Ptx} {P_{\text{T}}}

\newcommand{\pBS}{\textbf{p}_\text{BS}}

\newcommand{\pu}{\mathbf{p}_\text{u}}

\newcommand{\Thetak}{\Theta^{(k)}}

\newcommand{\sk}{s^{(k)}}

\acrodef{ABIM}{active backscattering intelligent metasurface}
\acrodef{ACK}{acknowledge}
\acrodef{AF}{array factor}
\acrodef{AOA}{angle-of-arrival}
\acrodef{AP}{access point}
\acrodef{AWGN}{additive white Gaussian noise}
\acrodef{BAV}{balanced antipodal Vivaldi}
\acrodef{BER}{bit error rate}
\acrodef{BS}{base station}
\acrodef{CDF}{cumulative distribution function}
\acrodef{CDL}{Clustered Delay Line}
\acrodef{CED}{channel excess delay}
\acrodef{ch.f.}{characteristic function}
\acrodef{CIR}{channel impulse response}
\acrodef{CPR}{channel pulse response}
\acrodef{CR}{channel response}
\acrodef{CRB}{Cram\'{e}r-Rao bound}
\acrodef{CRLB}{Cram\'{e}r-Rao lower bound}
\acrodef{CSI}{channel state information}
\acrodef{CW}{continuous wave}
\acrodef{DF}{detect \& forward}
\acrodef{DoF}{degrees-of-freedom}
\acrodef{DS}{delay spread}
\acrodef{i.i.d.}{independent, identically distributed}
\acrodef{ECDF}{empirical cumulative distribution function}
\acrodef{ED}{energy detector}
\acrodef{EIRP}{effective radiated isotropic power}
\acrodef{EM}{electromagnetic}
\acrodef{EVD}{eigen-value decomposition}
\acrodef{FCC}{Federal Communications Commission}
\acrodef{FIM}{Fisher information matrix}
\acrodef{FF}{far-field}
\acrodef{FMCW}{frequency modulated continuous wave}
\acrodef{GDOP}{geometric dilution of precision}
\acrodef{GLRT}{generalized likelihood ratio test}
\acrodef{GML}{generalized maximum likelihood}
\acrodef{GP}{Gaussian process}
\acrodef{HPBW}{half power beam width}
\acrodef{IIoT}{industrial Internet-of-things}
\acrodef{IF}{intermediate frequency}
\acrodef{IMF}{Ideal Matched Filter}
\acrodef{IMU}{inertial measurement unit}
\acrodef{INR}{interference-to-noise ratio}
\acrodef{IoT}{Internet-of-things}
\acrodef{IRS}{intelligent reconfigurable surface}
\acrodef{IR-UWB}{impulse radio UWB}
\acrodef{ISAC}{integrated sensing and communication}
\acrodef{ISI}{inter-symbol interference}
\acrodef{JF}{just forward}
\acrodef{LEO}{localization error outage}
\acrodef{LIS}{large intelligent surface}
\acrodef{LFM}{linear frequency modulated}
\acrodef{LOS}{line-of-sight}
\acrodef{LRT}{likelihood ratio test}
\acrodef{LS}{least squares}
\acrodef{MAC}{medium access control}
\acrodef{MIMO}{multiple-input multiple-output}
\acrodef{MIS}{medium intelligent surface}
\acrodef{MF}{matched filter}
\acrodef{ML}{Maximum Likelihood}
\acrodef{mMTC}{massive Machine Type Communication}
\acrodef{MMSE}{minimum-mean-square-error}
\acrodef{mmWave}{millimeter-wave}
\acrodef{MP}{metaprism}
\acrodef{MS}{mobile station}
\acrodef{MSE}{mean square error}
\acrodef{MUI}{multi-user interference}
\acrodef{MU-MIMO}{multi-user MIMO}
\acrodef{NBI}{narrowband interference}
\acrodef{NF}{near-field}
\acrodef{NL}{nonlinear}
\acrodef{NLOS}{non-line-of-sight}
\acrodef{OFDM}{orthogonal frequency division multiplexing}
\acrodef{PAM}{pulse amplitude modulation}
\acrodef{PEB}{position error bound}
\acrodef{PSD}{power spectral density}
\acrodef{PDF}{probability distribution function}
\acrodef{PDP}{power delay profile}
\acrodef{PHY}{physical layer}
\acrodef{ppm}{part-per-million}
\acrodef{PPM}{pulse position modulation}
\acrodef{PPP}{Poisson point process}
\acrodef{PR}{pseudo-random}
\acrodef{PRP}{pulse repetition period}
\acrodef{PSWF}{prolate spheroidal wave function}
\acrodef{RADAR}{RADAR}
\acrodef{RCS}{radar cross-section}
\acrodef{RF}{radiofrequency}
\acrodef{RFID}{radio frequency identification}
\acrodef{RIS}{reconfigurable intelligent surface}
\acrodef{RISs}{reconfigurable intelligent surfaces}
\acrodef{RRC}{root raised cosine}
\acrodef{RSS}{received signal strength}
\acrodef{RSSI}{received signal strength indicator}
\acrodef{RTT}{round-trip time}
\acrodef{RV}{random variable}
\acrodef{RMSE}{root mean square error}
\acrodef{SBS}{serial backward search}
\acrodef{SCR}{signal-to-clutter ratio}
\acrodef{SINR}{signal-to-interference-noise ratio}
\acrodef{SIR}{signal-to-interference ratio}
\acrodef{SIS}{small intelligent surface}
\acrodef{SoC}{System on Chip}
\acrodef{SoP}{System on Package}
\acrodef{SNR}{signal-to-noise ratio}
\acrodef{SPMF}{Single-Path Matched Filter}
\acrodef{S-V}{Saleh-Valenzuela}
\acrodef{SVD}{singular-value decomposition}
\acrodef{TDD}{time-division duplexing}
\acrodef{TDOA}{time difference-of-arrival}
\acrodef{TH}{time-hopping}
\acrodef{TNR}{threshold-to-noise ratio}
\acrodef{TOA}{time-of-arrival}
\acrodef{TOF}{time-of-flight}
\acrodef{TR}{time-reversal}
\acrodef{UE}{user equipment}
\acrodef{UHF}{ultra-high frequency}
\acrodef{ULA}{uniform linear array}
\acrodef{UWB}{ultrawide bandwidth}
\acrodef{WBI}{wideband interference}
\acrodef{WPAN}{wireless personal area networks}
\acrodef{WSN}{wireless sensor network}
\acrodef{WSR}{wireless sensor radar}
\acrodef{WSS}{wide-sense stationary}
\acrodef{WWLB}{Weiss-Weinstein lower bound}

\begin{document}

\title{NLOS Localization\\ Exploiting Frequency-selective Metasurfaces}

\author{~Marina~Lotti,~\IEEEmembership{Graduate~Student~Member,~IEEE},~Giacomo~Calesini,\\~Davide~Dardari,~\IEEEmembership{Senior Member,~IEEE}
\thanks{Manuscript received July 12, 2023; revised MM DD, AAAA.\\
This work was supported, in part, by the European Union under the Italian National Recovery and Resilience Plan
(NRRP) of NextGenerationEU, partnership on “Telecommunications of the Future” (PE00000001 - program “RESTART”), and by the EU Horizon project TIMES (Grant no. 101096307).

The authors are with the Department of Electrical, Electronic and Information Engineering ``Guglielmo Marconi”, University of Bologna, 47521 Cesena Campus, Italy, and also with the WiLAB-CNIT, 40136 Bologna, Italy (e-mail: marina.lotti2@unibo.it; davide.dardari@unibo.it; giacomo.calesini@studio.unibo.it).
}
}


\maketitle

\begin{abstract}

This paper introduces a new approach to localize user devices located in non-line-of-sight (NLOS) areas using a passive, non-reconfigurable, and frequency-selective metasurface called metaprism. By analyzing the spatial filtering of subcarriers in the orthogonal frequency division multiplexing (OFDM) signal transmitted by each user device, the base station can estimate the device's angle of view, distance, and subsequently its position. 
Two different criteria are proposed for designing the frequency response of the metaprism, depending on whether the users operate in the far-field or near-field region of the metaprism.
Simulation results in the millimeter-wave band demonstrate that the system can achieve an accuracy of less than 2 degrees in angle estimation and in the order of decimeters in position estimation.

\end{abstract}


\section{Introduction}
\IEEEPARstart{I}{n} the next-generation wireless networks, localization will be one of the key applications \cite{DeLetAll:2021}. To be able to locate users in hostile environments such as \ac{IIoT}, characterized by frequent signal blockage when working at very high frequency, one possible solution to avoid the deployment of a very large number of \acp{BS} is the utilization of \acp{RIS}. These surfaces 
are capable of generating anomalous reflections of the signals to establish virtual \ac{LOS} conditions \cite{CisAhmSezWym:21,DarDecGueGiu:21}. 
However, \acp{RIS} have the disadvantage of requiring reconfiguration, energy supply, the estimation of the \ac{CSI}, and dedicated control channels, thus making them not always a viable and cheap solution. 

In this work, we instead adopt a completely passive, non-reconfigurable, and frequency-selective metasurface, called \textit{metaprism}, recently defined and considered in \cite{DarMas:21} to extend the communication coverage without the disadvantages of \acp{RIS}. 
The exploitation of frequency-selectivity effects of antennas/metasurfaces is not new in the literature.
For instance, various types of frequency beam-scanning array antennas operating at high frequencies have been developed, as done in \cite{SarJamVahEas:18}. These antennas can be utilized, for example, to estimate the direction of arrival in azimuth and elevation, as demonstrated in \cite{OrtKwiPoh:19}. In \cite{MenWanWanZheLiMaQu:22}, the authors employed a frequency-selective metasurface as the target to be detected and localized by a bistatic and monostatic radar.

Differently from previous literature, in our work, the metaprism is neither the target to be localized nor the antenna at the \ac{BS}/radar used to localize a target/user. Here the metaprism is used as a passive reflector to assist the \ac{BS} in the task of localizing a user located in \ac{NLOS} condition in a similar way as done using \acp{RIS} but without the need for power supply, \ac{CSI} estimation, reconfiguration, and signaling overhead. In particular, we propose two different criteria for designing the frequency response of the metaprism, depending on whether the users operate in the \ac{FF} or \ac{NF} region of the metaprism, and a procedure to estimate the user's angle of view and distance by exploiting  the spatial filtering of subcarriers in the \ac{OFDM} signal transmitted by each user. Numerical results corroborate the validity of the approach.
 
\begin{figure}
    \centering
    \includegraphics[width=0.5\linewidth]{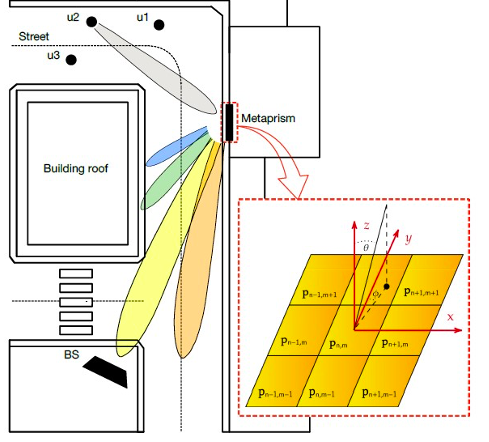}
    \caption{Typical scenario in which the signal transmitted by a user is reflected by the metaprism towards the base station.
    Each subcarrier component is reflected toward a different  angle.}
    \label{fig: scenario city}
\end{figure}

\section{System Model and Metaprim Design}
Referring to Fig.~\ref{fig: scenario city}, we suppose a certain number of users are located in a region that is in \ac{NLOS} with respect to the \ac{BS} located at coordinates $\pBS=(d_{\text{BS}}, \Theta_{\text{BS}})$, where $d_{\text{BS}}$ and $\Theta_{\text{BS}}$ are, respectively, the distance and the 3D angle of the \ac{BS} with respect to the center $\mathbf{p}_0 = (0,0,0)$ of the metaprism, being $\Theta_{\text{BS}} = \left( \theta_{\text{BS}}, \phi_{\text{BS}} \right)$, with $\theta_\text{BS}$ and  $\phi_{\text{BS}}$ representing  the  azimuth and elevation angles.  We assume that both the \ac{BS} and the users are in \ac{LOS} with reference to the metaprism. To be localized, the users send pilot symbols using an \ac{OFDM} modulation scheme in the available $K$ subcarriers according to any multiple access protocol to avoid inter-user interference. The analysis of multiple access schemes is out of the scope of this paper then, in the following, we consider the localization of a generic single user located at position $\pu=(d_{\text{u}}, \Theta_{\text{u}})$, where  $\Theta_{\text{u}} = \left( \theta_{\text{u}}, \phi_{\text{u}} \right)$. 
Denote by $f_0$ the carrier frequency, $\lambda$ the corresponding wavelength, and by $f_k$ the $k$th subcarrier frequency within the bandwidth $W$.  

We suppose  the metasurface lays in the $x-y$ plane consisting  of $N \times M $ cells whose size is $d_x \times d_y$, where $d_x = d_y \approx \lambda/2$. Denote by  $\mathbf{p}_{n,m} = \{x_n, y_m,0\}$, the position of the $nm$th cell, where $x_n = n \, d_x - N\,  d_x /2 $, with $n = 0,1,\dots N-1$, and $y_m = m\,  d_y - M \, d_y /2 $, with $m = 0,1,\dots M-1$. 
We consider the $nm$th cell of the metaprism  being characterized by the following frequency-selective reflection coefficient \cite{DarMas:21} 
\begin{equation}
    r_{nm}(f)  =  e^{j \Psi_{nm}(f)}
\end{equation}
where $\Psi_{nm}(f)$ is the frequency-dependent reflection phase profile. 
According to the model in \cite{DarMas:21}, $\Psi_{nm}(f)$  can be designed such that it  exhibits a linear behavior with the frequency $f$, i.e.,
\begin{equation}
    \Psi_{nm}(f) = \alp \cdot  (f - f_0 ) 
\label{eq:pp_pos}
\end{equation}
where $\alp$ is  a cell-dependent coefficient. According to \eqref{eq:pp_pos}, the frequency-dependent phase profile of the metaprism depends on the design of coefficients $\alp$. As it will be shown, such a design allows the \ac{BS} to infer the position of the user starting from the reception of the reflected signal. In the following, we propose two design criteria tailored to the operating condition of the system: i) users and \ac{BS} located in the \ac{FF} region of the metaprism; ii) users and \ac{BS} located in the  \ac{NF} region.

\subsection{FF scenario: Beamsteering design (BD)}
When both the user and the \ac{BS} are in the \ac{LOS} \ac{FF}-region with respect to the metaprism, it is possible to discriminate only the angle of view of the user but not its distance from the metaprism. This can be done by properly designing the coefficients $\{\alpha_{nm}\}$ of the cells so that subcarrier-dependent beamsteering is obtained. In this way, the angle of the user is  directly linked to the subcarriers and it can be estimated from their observations, as it will be explained in the next section. 
The proposed criterium for designing the coefficients $\{\alpha_{nm}\}$ in \eqref{eq:pp_pos} is expressed by 
\begin{equation}
    \alpha_{nm} = a_0\,x_n + b_0\,y_m \, .
    \label{eq:alpha_FF}
\end{equation}
According to this model,  in the following referred to as  \textit{beamsteering design (BD)}, $\alpha_{nm}$ increases incrementally based on the cell's coordinates, and $a_0$ and $b_0$ are two constants that need to be appropriately determined. 

\begin{figure}
    \centering
    \includegraphics[trim=1cm 1cm 1cm 1cm, width=0.8\linewidth]{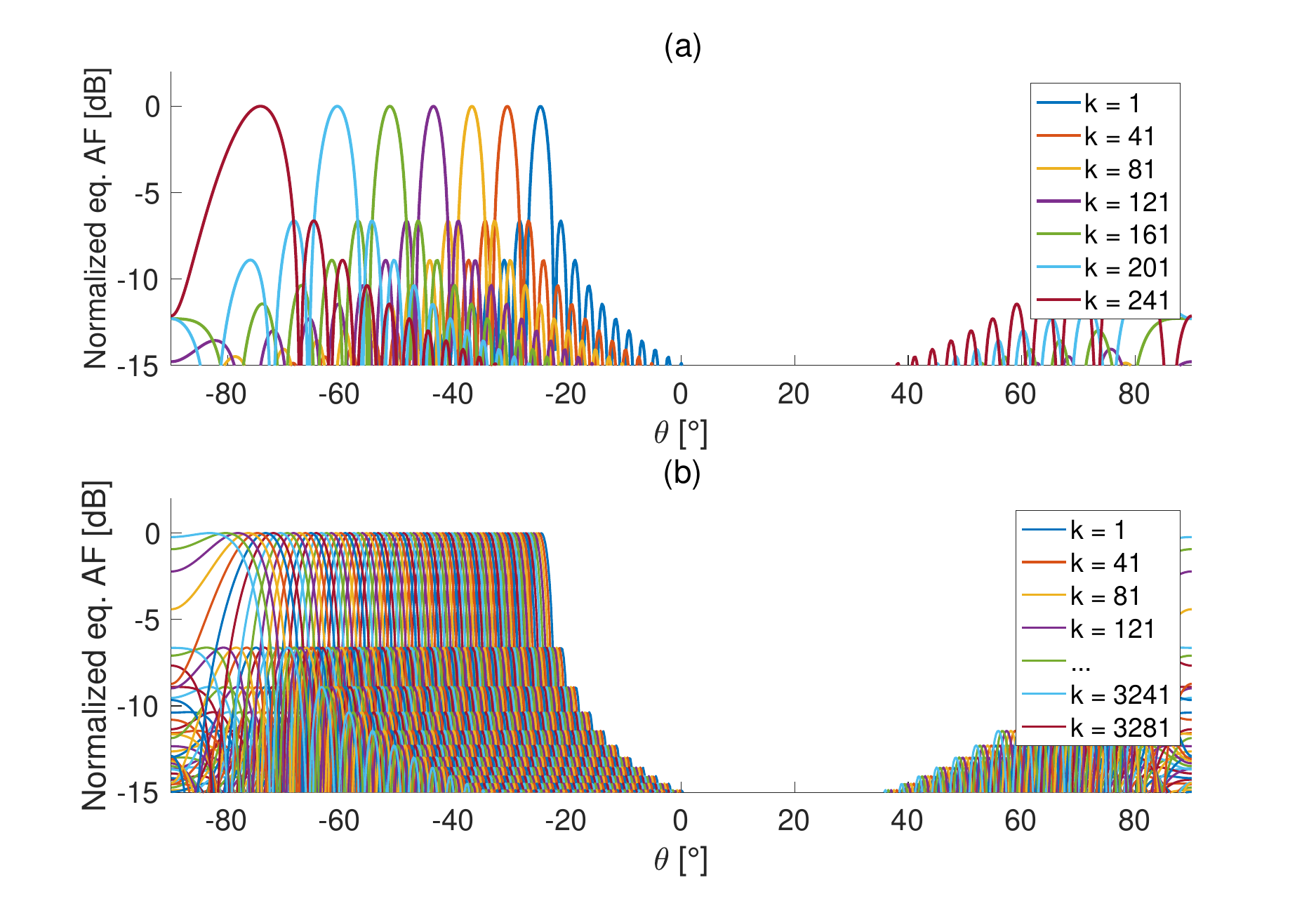}
    \caption{Normalized equivalent AF for $K=256$ (a) and $K=3300$ (b). }
    \label{fig: af plot}
\end{figure}

It is worth noticing that the user-metaprism subsystem is equivalent to a frequency-selective reflectarray antenna characterized by  the following subcarrier-dependent \ac{AF} 
 \begin{align}
    AF^{(k)}(\Theta) =& \sum_{n=0}^{N-1} \sum_{m=0}^{M-1}\,\exp \left(j\frac{2\pi\,n\,d_x}{\lambda}\,\left(u_x(\Theta) + u_x(\Theta_{\text{u}})\right) \notag \right.\\
    &+\left. j\frac{2\pi\,m\,d_y}{\lambda}\,\left(u_y(\Theta) + u_y(\Theta_{\text{u}})\right) + \jmath \Psi_{nm}(f_k)  \right)
    \label{eq:AF}
  \end{align}
  where we have defined the quantities $ u_x (\Theta) = \sin(\theta) \cos(\phi)$ and $ u_y(\Theta) = \sin(\theta) \sin(\phi)$.
The \ac{AF}  for a certain direction $\Theta$ indicates the intensity with which the $k$th subcarrier of the signal incident on the metaprism generated by the user is reflected in that direction. 
From \eqref{eq:pp_pos},  \eqref{eq:alpha_FF} and by equating to zero the total phase shift in \eqref{eq:AF}, it is easy to show that the  angle of reflection $\Thetak = \left( \theta^{(k)}, \phi^{(k)}\right)$ of the $k$th subcarrier at frequency $f_k$ of the incident OFDM signal is given by
\begin{align}
    & u_x \left (\Thetak \right ) = -u_x (\Theta_{\text{u}} ) - \frac{ a_0 \lambda}{2 \pi } \left( f_k - f_0\right) \notag \\
    & u_y \left  (\Thetak \right )  = -u_y (\Theta_{\text{u}} ) - \frac{b_0 \lambda}{2 \pi } \left( f_k - f_0\right)  \, .
    \label{eq:reflection_direction}
\end{align}
Now, suppose we want to design the metaprism so that it reflects the highest subcarrier of the user-generated signal ($k=K$) with angle $\theta^{(K)} = - \theta_{\text{u}} - \theta_m$, for some angle $\theta_m$, and $\phi^{(K)}=0$.  From \eqref{eq:reflection_direction}, by setting $k = K$, the  coefficients $a_0$ and $b_0$ become
\begin{equation} \label{eq:a0}
a_0 = {-4 \pi}/{ \lambda W} \left( -\sin\left( \theta_{\text{u}} + \theta_m\right) + \sin\left( \theta_{\text{u}} \right) \right)
\end{equation}
and $b_0 = 0$. With these values, the other subcarriers are reflected with different increasing angles in the range $\theta^{(k)} \in [ \theta^{(K)},\theta^{(1)}]$, where $\theta^{(1)}$ can be obtained from \eqref{eq:reflection_direction} with $k = 1$, around the Snell’s angle $-\theta_\text{u}$. 
Thanks to the reciprocity of the metaprism, by exchanging $\theta_{\text{u}}$ with $\theta_{\text{BS}}$,  the same design corresponds to a metaprism that reflects the different subcarriers  toward the users depending on the \ac{BS}'s angle $\theta_{\text{BS}}$.
Therefore, according to the angle $\theta_m$, it is possible to design the desired angular span necessary to cover a specific \ac{NLOS} area of interest. 
Fig.~\ref{fig: af plot} shows an example of the normalized \ac{AF} in dB  obtained from \eqref{eq:AF}, with a metaprism  designed using \eqref{eq:a0} with $\theta_m=40^\circ$ so that the main lobe moves from $-25^\circ$ to $-85^\circ$, considering a $50\times 50$ cells metaprism, for the case of $K = 256$ and $K = 3300$ (the other parameters are those used in the numerical results section). 
For a fixed user's position (incident angle),  it can be observed that as the number of subcarriers increases from $K=256$ to $K=3300$, the peaks of the equivalent \ac{AF}  become increasingly dense thus increasing the probability that the \ac{BS} receives a sufficient amount of energy at least in one subcarrier. Notably, if the angle of the user is changed, a different subcarrier will be directed toward the \ac{BS}. This is the phenomenon that will be exploited by the position estimator described in Sec. \ref{Sec:PosEst}.

\subsection{NF  Scenario: Random design (RD)}
\label{NF}
In \ac{NF} channel conditions, the plane wave propagation approximation is no longer valid and the BD criterium of the previous section cannot be applied 
because the cells of the metaprism do not share anymore the same angle of arrival/departure \cite{DarDecGueGiu:21}. 
Therefore, the design of  coefficients $\{\alpha_{nm}\}$ in \eqref{eq:pp_pos} needs to be re-adapted to the \ac{NF} case. 
One option  would be designing $\{\alpha_{nm}\}$ to obtain a focusing effect, as done in \cite{DarMas:21}. Unfortunately, this leads to a model with 3 parameters instead of 2 in \eqref{eq:alpha_FF} that does not give the necessary flexibility for localization purposes. 
For this reason, we propose to draw $\{\alpha_{nm}\}$ randomly from a uniform distribution (\emph{Random design - RD}) as $\alpha_{nm} \sim {\cal UD} (0,10^{-6})$,
within the same range of values of the beamsteering design. 
The rationale behind the RD is based on the idea that assigning random values to the metaprism's coefficients guarantees a unique or nearly unique frequency profile for different user's positions. 
By considering a large number of subcarriers and leveraging the randomness, the probability that two different positions have the same frequency profile is expected to decrease significantly. Unlike the BD approach, where the signal is reflected coherently towards a specific  direction thus ensuring a certain array gain, the random design does not exploit the array gain, and the overall link budget tends to degrade. 
\section{Position Estimation } \label{Sec:PosEst}
Considering a Ricean channel model with Rice factor $\kappa$, the signal received by the \ac{BS} at the $k$th subcarrier  can be written as 
\begin{equation}
    y_k = s_k(\pu) + v_k + n_k
\end{equation}
for $k=1, 2, \ldots, K$, where $s_k(\pu)$ is the specular component of the useful signal at the $k$th subcarrier which depends on the user's position $\pu$, whereas 
$n_k \sim \mathcal{CN}(0,\,\sigma^{2})\,$  is the thermal noise sample, being $\sigma^{2} = N_0 \Delta f$, $N_0$ the one-side noise power spectral density, and $\Delta f$  the carrier spacing $\Delta f=W/K$. The term $v_k$ represents the diffuse component of the received signal and it is modeled as a complex Gaussian random variable, i.e.,     
 $v_k \sim \mathcal{CN}(0,\,\sigma^{2}_v)\,$. 
The useful specular component of the received signal  $s_k(\pu)$ is given by
\begin{equation}
    s_k(\pu) =\sum_{n=0}^{N-1} \sum_{m=0}^{M-1} h_{mn}^{(k)} (\pu)\, r_{mn}^{(k)}\, g_{mn}^{(k)} (\pBS)  
    \times \sqrt{P_k \frac{\kappa}{1 +\kappa}} x_k 
    \label{eq:sk}
\end{equation}

where $r_{mn}^{(k)}=r_{mn}(f_k)$, $P_k = \Ptx/K$ is the transmitted power allocated to each subcarrier, $\Ptx$ the total transmitted power, and $x_k$ is the pilot symbol we set to one without loss of generality. $h_{mn}^{(k)} (\pu)$ and $g_{mn} ^{(k)}(\pBS)$ are the channel gains at frequency $f_k$ between the user and the  $nm$th cell of the metaprism and between the $nm$th  cell and the \ac{BS}, respectively, given by
\begin{align}
    & h_{mn}^{(k)}(\pu)=\frac{ \sqrt{G_\text{u}}\,\lambda}{4\pi|\pu-p_{mn}|} \exp \left(-j \frac{2\pi f_k}{c} |\pu-p_{mn}|\right) \\
    & g_{mn}^{(k)}(\pBS)=\frac{ \sqrt{G_\text{BS}}\,\lambda}{4\pi|\pBS-p_{mn}|} \exp \left(-j \frac{2\pi f_k}{c} |\pBS-p_{mn}|\right) 
    \label{eq:channelss}
\end{align}
being $c$ the speed of light, $G_\text{u}$ and $G_\text{BS}$ the antenna gains of the user and \ac{BS}, respectively. 
According to the Ricean fading model, it is  $  \sigma^{2}_v = {|A_k|^2 \Ptx}/{(1+\kappa)}$,  having defined
\begin{equation} 
A_k = \sum_{n=0}^{N-1} \sum_{m=0}^{N-1} h_{mn}^{(k)}(\pu) \, r_{mn}^{(k)}  \, g_{mn}^{(k)} (\pBS) \, .
\end{equation}  
Since the profile $\{y_k\}$, for $k=1, 2, \ldots, K$, of the received signal  is a function of the user's position $\pu$, then it is possible to estimate  $\pu$ from the observation of the $y_k$'s. 
In this work, we exploit the \ac{ML} estimator, which can be expressed as 
\begin{equation}
   \widehat{\mathbf{p}}_{\text{u}}= \underset{\pu}{\arg\max} \ \sum_{k=0}^{K-1} Re(y_k\, s_k^*(\pu)) \, .
   \label{eq: stimatore}
\end{equation} 
The \ac{ML}   in \eqref{eq: stimatore} corresponds to  performing ``fingerprinting'' localization: the estimator compares the test profiles $\{s_k(\pu)\}$, computed in all the test locations of the area of interest, with the profile of the received signal $\{y_k\}$, and chooses the position of the profile test pattern that most closely resembles that of the received signal. Obviously, the accuracy of the \ac{ML} is strictly affected by the granularity of the  test grid.

\section{Numerical Results}\label{Sec:results}
\begin{figure}[t]
    \centering
    \includegraphics[trim=1cm 1cm 1cm 1cm, width=0.6\linewidth]{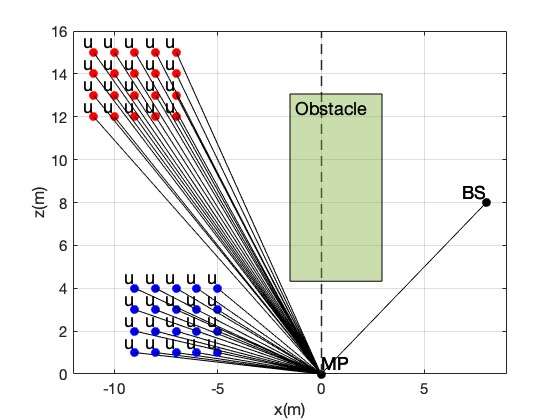}
    \caption{Users positions in scenario A (blue dots) and scenario B (red dots). }
    \label{fig: scenario}
    \vspace{-0.4cm}
\end{figure}
\begin{figure}
    \centering
    \includegraphics[trim=1cm 1cm 1cm 1cm, width=0.6\linewidth]{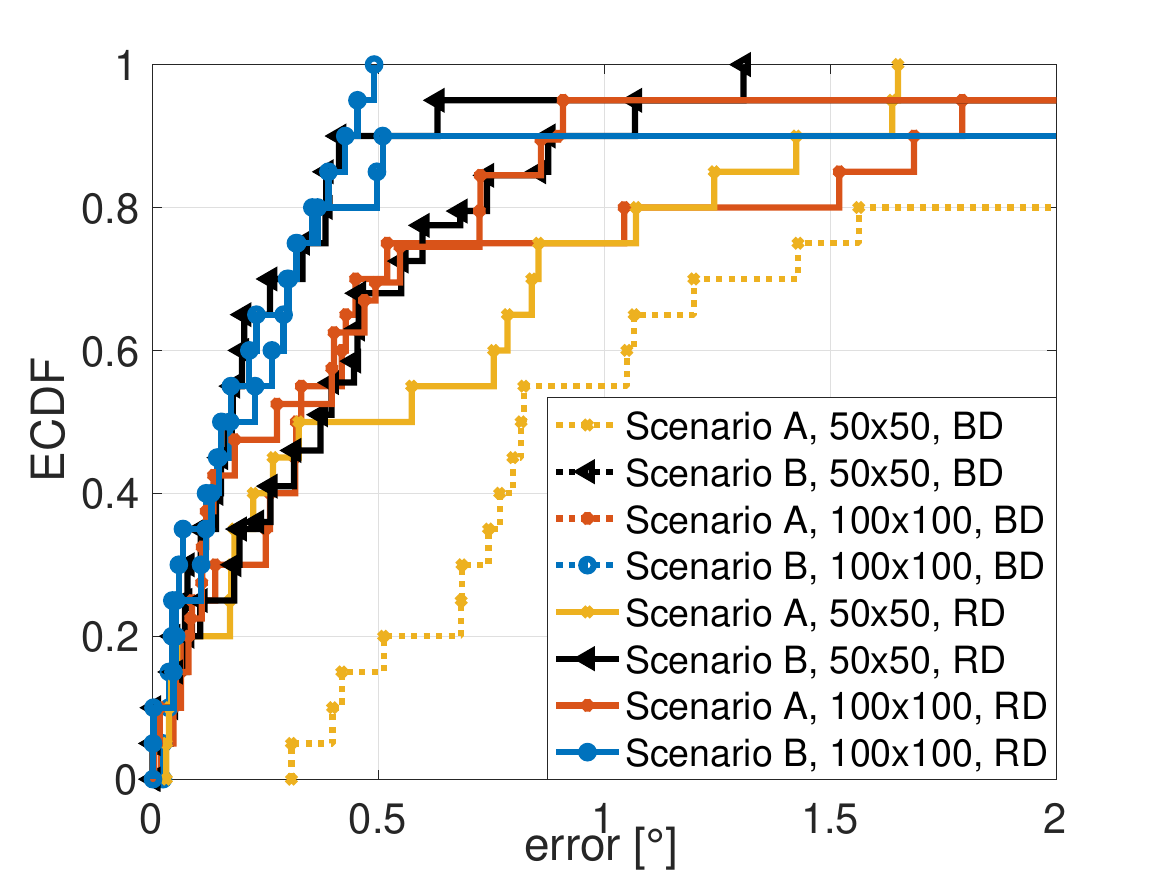}
    \caption{Angular ECDF varying the metasurface's size ($N \times M$). \\ BD (beamsteering design); RD (random design). No fading. }
    \label{fig:Ang BD_RD}
\end{figure}
\begin{figure}[t]
    \centering
    \includegraphics[trim=1cm 1cm 1cm 1cm, width=0.6\linewidth]{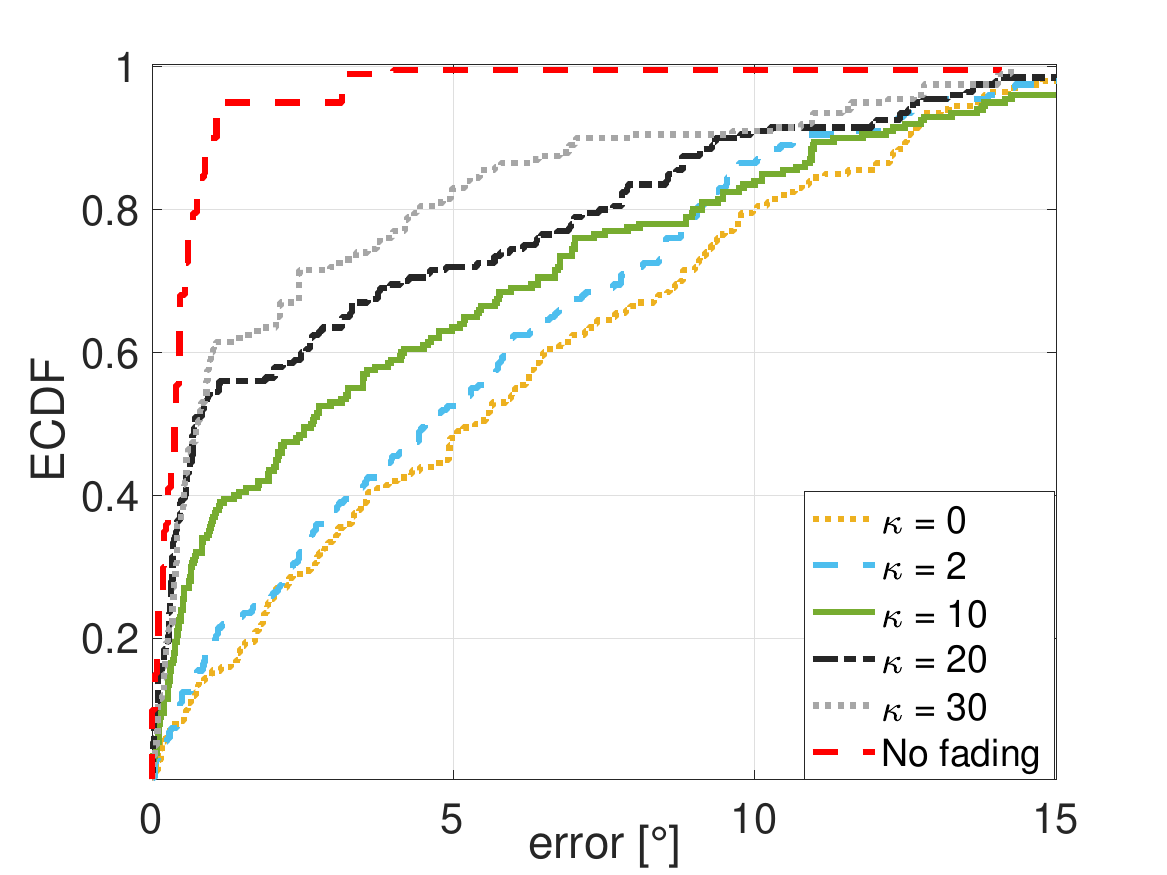}
    \caption{Angular ECDF by varying the Rice Factor $\kappa$ in scenario B. }
    \label{fig:ANG_BD_fading}
\end{figure}
\begin{figure}
    \centering
    \includegraphics[trim=1cm 1cm 1cm 1cm, width=0.6\linewidth]{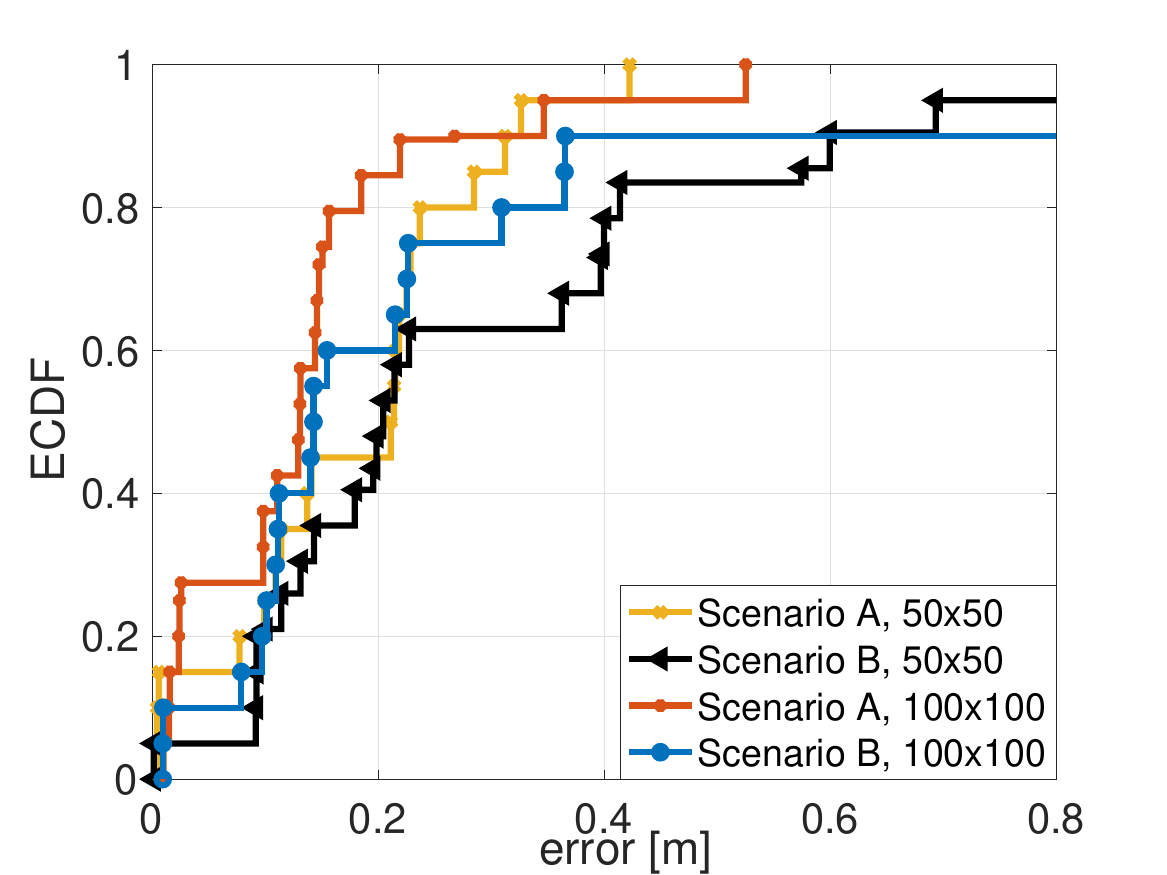}
    \caption{Position estimation error ECDF, random design, no fading. }
    \label{fig:Dist_Pos_RD}
\end{figure}

In this section, we illustrate some simulation examples addressed to validate the proposed metasurface-aided localization schemes. 
The two \ac{NLOS} scenarios considered in the simulations are sketched in Fig.~\ref{fig: scenario} in which the red dots represent the user's positions tested in the \ac{FF} region, while the blue dots represent the user's positions tested  in the \ac{NF} one.  It is important to note that the Fraunhofer distance varies depending on the dimensions of the metaprism considered, as $d_{F}=2D^2/\lambda$, where $D$ is the maximum size of the metaprism. To avoid confusion, we will refer to the scenario with blue dots as \textit{scenario A} and the scenario with the red dots as \textit{scenario B}.
The value of the parameters resembles those of the mmWave 5G-NR wireless systems, i.e., $f_0=28\,$GHz, $P_T =20\,$dBm, $G_\text{BS} = G_\text{u} = 6\,$ dB, bandwidth $W=198\,$MHz, $K= 3300$, \ac{BS}'s position $\textbf{p}_\text{BS}=\{8,0,8\}$ m, receiver's noise figure $F_\text{noise}=$ 3 dB. The test grid step is equal to $0.1\,$m. 
For each system configuration and metaprism design criterium,  a 200-iteration Monte Carlo simulation has been performed.
The effect of the metaprism's size is investigated in 
Fig.~\ref{fig:Ang BD_RD} where the \acp{ECDF} of the angle estimation error is reported. 
As expected, by increasing the size from $N\times M = 50\times50$ ($26.7 \times 26.7 \, \text{cm}^2$) to $100\times100$ ($53.5 \times 53.5\, \text{cm}^2$), also the performance improves thanks to better link budget and angular resolution. For the BD, the estimation error is less for users in scenario B. This is due to the fact that the BD assumes the users are located in the \ac{FF} region (scenario B), while for the RD, this difference is not so evident.
The impact of fading can be analyzed in Fig.~\ref{fig:ANG_BD_fading} for different values of the Rice factor $\kappa$. 
For the sake of space, only the case of scenario B for BD is reported, but the trend is the same for scenario A and RD. 
As expected, with $\kappa = 0$ (Rayleigh fading), where the useful signal $\sk(\pu)$ is null and only the noisy components are present, the estimation is not possible. In fact, the corresponding curve refers to the uncertainty associated with the spanned area whose size is a priori known. Vice versa, the curves related to higher values of $\kappa$ indicate that the proposed scheme is effective in strong Ricean fading, which is typical at millimeter waves or higher frequencies.     
The RD criterium opens the possibility to estimate not only the angle, as in the BD, but also the user's position, i.e., the angle and distance.
This can be observed in Fig.~\ref{fig:Dist_Pos_RD} where the \acp{ECDF} of the position estimation error is reported for different configurations. 
The position estimation has a maximum error of less than $22\,$cm for scenario A and $35\,$cm for scenario B in $90\%$ of locations. 
%
%
%
\section{Conclusion}
In this work, we have proposed an approach  exploiting a passive, non-reconfigurable, and frequency-selective metasurface (metaprism) to localize a user in  \ac{NLOS} condition.
The numerical results obtained at millimeter waves demonstrate that the system can achieve an angle estimation error of about 2$^\circ$ and a position estimation error lower than  $40\,$cm in 90$\%$ of the cases considered. 
%

\bibliographystyle{IEEEtran}
\bibliography{biblio}

\end{document}